\documentclass[]{article}
\usepackage[utf8]{inputenc}
\usepackage{amsmath}
\usepackage[pdftex]{graphicx}
\usepackage{amscd}
\usepackage{calrsfs}
\usepackage{slashed}
\usepackage{amsfonts}
\usepackage{authblk}
\usepackage{placeins, float}
\usepackage[letterpaper, portrait]{geometry}

\title{Antipodal identification in the Schwarzschild spacetime}
\author{Miguel Socolovsky}
\affil{Instituto de Ciencias Nucleares, Universidad Nacional Aut\'onoma de M\'exico, Cd. Universitaria, 04510, Ciudad de M\'exico, M\'exico\\socolovs@nucleares.unam.mx}

\providecommand{\keywords}[1]
{
	\small	
	\textbf{Keywords:} #1
}
\begin{document}
\date{}
\maketitle

\begin{abstract}
Through a Möbius transformation, we study aspects like topology, ligth cones, horizons, curvature singularity, lines of constant Schwarzschild coordinates $r$ and $t$, null geodesics, and transformed metric, of the spacetime $(SKS/2)^\prime$ that results from: i) the antipode identification in the Schwarzschild-Kruskal-Szekeres ($SKS$) spacetime, and ii) the suppression of the consequent conical singularity. In particular, one obtains a non simply-connected topology: $(SKS/2)^\prime\cong \mathbb{R}^{2*}\times S^2$ and, as expected, bending light cones. 
\end{abstract}

\keywords{antipodal identification; Schwarzschild spacetime}

\section{Introduction}
In 1965, Rindler suggested the possibility of the antipode identification of coordinates $(V,U)\equiv (-V,-U)$ in the maximal analytic extension of the Schwarzschild metric, the Schwarzschild-Kruskal-Szekeres ($SKS$) spacetime. Later, Sanchez and Whiting (1987) and more recently 't Hooft (2018) used it (or a variation of it) for the study of quantum field theory in black holes. The identification allows to obtain $SKS/2$, closer than $SKS$ to the physical spacetime associated with the total collapse of a spherically symmetric star: both the mirror image of ``our" asymptotically flat region, and the white hole region together with its associated past singularity $s_p$, dissappear. The conical singularity appearing as a consequence of this identification must be excluded from the spacetime, leading to the space $(SKS/2)^\prime$. Inserting its $(V,U)$ part in the upper part of a complex half plane, a Möbius-type complex transformation $\phi$ exhibits $(SKS/2)^\prime$ with the topology $(S^2_1\setminus (\{N,S\}\cup\phi(s_f)\cup int(\phi(s_f))))\times S^2\cong \mathbb{R}^{2*}\times S^2$, where $N$ and $S$ are the north and south poles of the unit 2-sphere $S^2_1$ and $s_f$ is the future singularity. Precisely due to the suppression of the conical singularity, its boundary, image of the past horizon -which dissappears in a physical collapse- can not be associated to a closed causal (null) curve. Also, the same suppression produces a non simply-connected space, since $\pi_1((SKS/2)^\prime)\cong\mathbb{Z}$. 
\section{Antipode identification}
The maximal analytic extension of the Schwarzschild metric (with coordinates $(t,r,\theta,\varphi)$, $t\in(-\infty,+\infty)$, $r>0$, $\theta\in(0,\pi)$, $\varphi\in[0,2\pi)$), is that of Kruskal-Szekeres (1960) with coordinates $(V,U,\theta,\varphi)$ ($V\in(-\infty,+\infty)$: temporal, $U\in(-\infty,+\infty)$: spatial, $\theta,\varphi$ as in Schwarzschild, angular coordinates). The $V/U$ part is given by the following diagram:

\begin{figure}[h!]
	\centering
	\includegraphics[width=\linewidth]{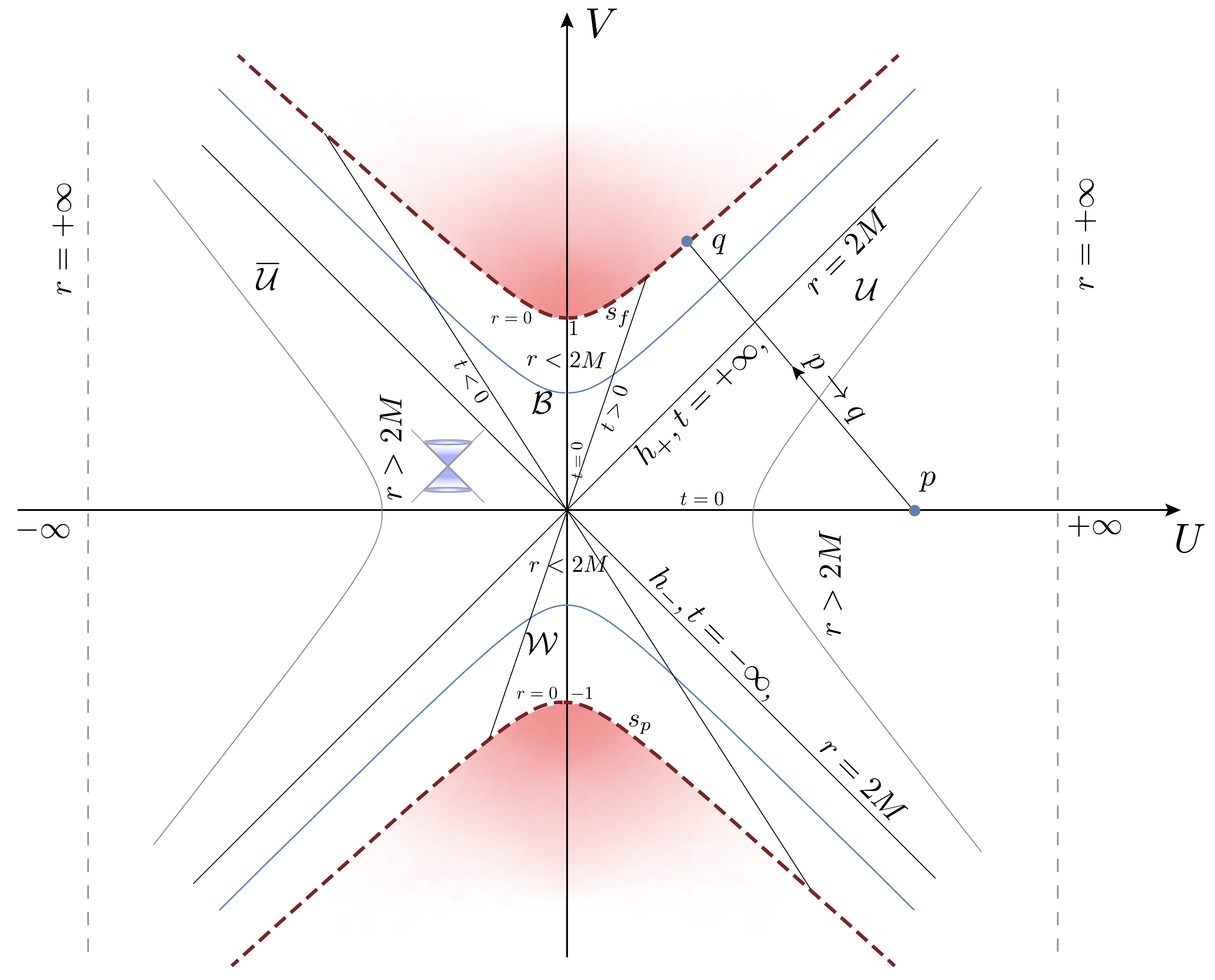}
	\caption{$SKS$ solution}
	\label{fig:1}
\end{figure}

In Fig. 1: $\cal{U}$: ``our" universe; $\cal{\bar{U}}$: anti-universe; $\cal{B}$: black hole; $\cal{W}$: white hole; dashed red lines: future and past singularities $s_f$ and $s_p$, respectively given by the hyperboles $V=\pm\sqrt{1+U^2}$, which are asymptotic to the future and past horizons $h_+$, $h_-$; $h_+$, $h_-$ and $s_f$, $s_p$ consist of points where $r=2M$ and $r=0$ respectively; the remaining hyperbole filling the $V/U$ plane correspond to fixed values of $r$, and the lines through the origin correspond to fixed values of $t$; each point has associated with it a 2-sphere $S^2=S^2_r$ of radius $r$, $S^2(r,\theta,\varphi)$; the light cones are at $\pm 45^\circ$ everywhere: particle trajectories in their interior are timelike and on their boundaries are null (light rays) ; $s_f$, $s_p$ and the shaded regions do not belong to the $SKS$ spacetime, which turns out to be non compact, 1-connected, globally hyperbolic (it has a global Cauchy surface) and geodesically incomplete; $M$ is the gravitating mass; since $V,U\in(-\infty,+\infty)$, then, topologically, 
\begin{equation}
SKS\cong\mathbb{R}^2\times S^2.
\end{equation}   
It can be shown that the ``lines" $h_\pm$ (together with the corresponding $S^2_{2M}=S^2(r=2M,\theta,\varphi)$) are null hypersurfaces; for the homotopy groups,
\begin{equation}
\pi_k(SKS)\cong\pi_k(\mathbb{R}^2\times S^2)\cong\pi_k(S^2)
\end{equation}
(0 for $k$=1, $\mathbb{Z}$ for $k$=2,3, $\mathbb{Z}_2$ for $k$=4, etc.); $\cal{U}$ and $\cal{\bar{U}}$ are asymptotically flat at $r\to+\infty$ (Minkowski spacetime). 

\

The $SKS$ metric is
\begin{equation}
ds^2={{32M^3}\over{r}}e^{r/{2M}}(dV^2-dU^2)-r^2(d\theta^2+sin^2\theta d\varphi^2),
\end{equation}
where the relations between the $(r,t)$ and $(V,U)$ is given by 
\begin{equation}
(1-{{r}\over{2M}})e^{r/2M}=V^2-U^2
\end{equation}
which gives $r=r(V,U)$ through the Lambert $W$ function defined via 
$\mu e^\mu=\nu\Rightarrow \mu=W(\nu)$, (Lambert, 1758), and 
\begin{equation}
{{t}\over{4M}}=Th^{-1}(({{V}\over{U}})^{sg(r/2M-1)\times 1}), \ r\neq 2M,
\end{equation}
with $t=\pm\infty$ at the horizons $V=\pm U$.

\

$ds^2$ has the symmetry $PT$ with $P$ the spatial inversion $U\to-U$ and $T$ the time inversion $V\to-V$, so that $PT:(V,U)\to (-V,-U)$. Since $(PT)^2=Id$, the associated symmetry group is $(Id,PT)\cong\mathbb{Z}_2$. This permits the antipode (through the origin $(V,U)=(0,0)$) identification $(-V,-U)\equiv(V,U)$ and allows to remain with only the ``half" spacetime in the Fig. 1, which we call $SKS/2$. This is illustrated in Fig. 2.

\begin{figure}[H]
	\centering
	\includegraphics[width=.5\linewidth]{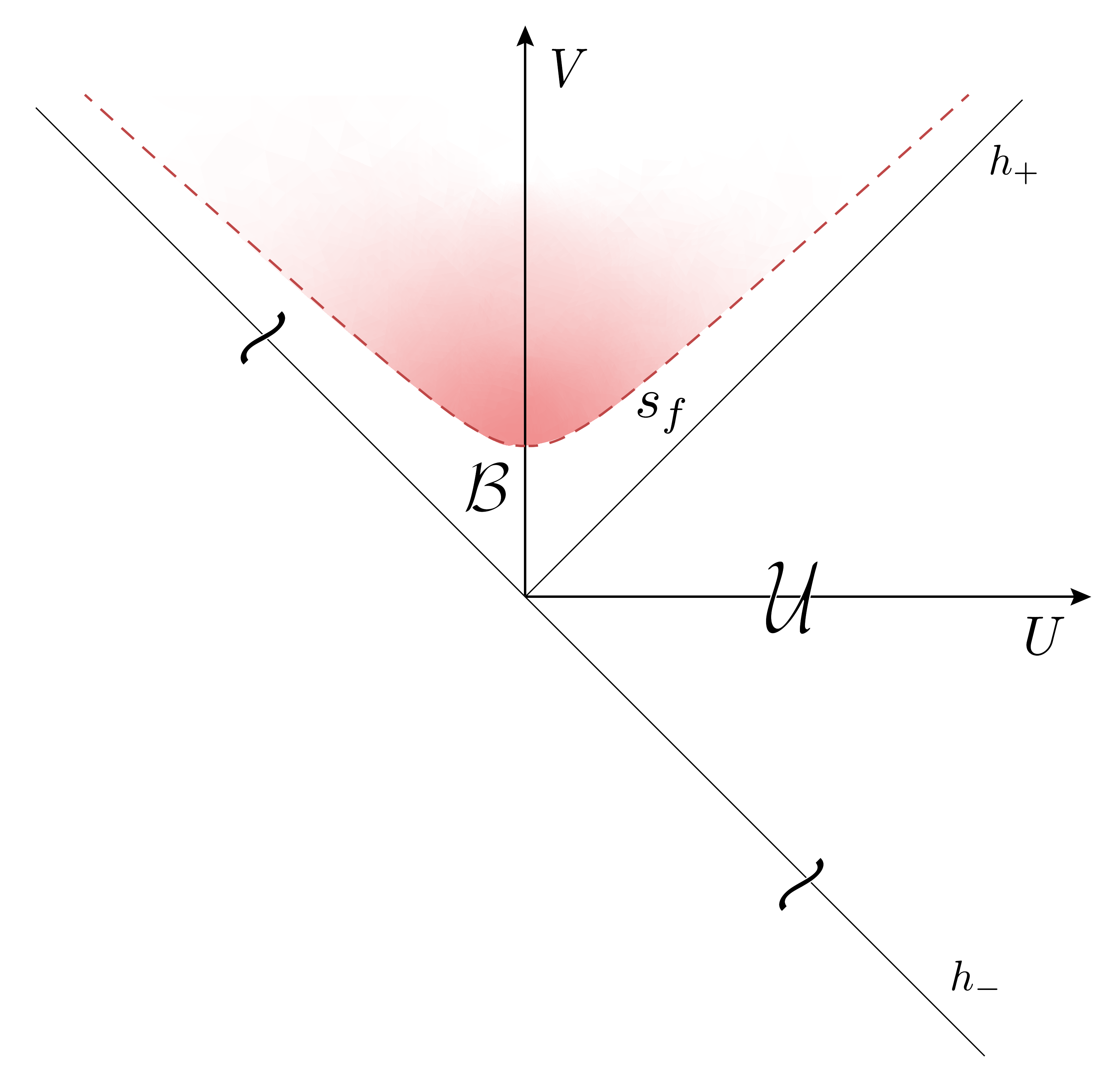}
	\caption{$SKS/2$}
	\label{fig:2}
\end{figure}

If $S^2(r,\theta,\varphi)$ were identified with $S^2(r,\pi-\theta,\varphi+\pi)$, then at each $r$ one should have the projective space $S^2/\mathbb{Z}_2\cong \mathbb{R}^3/\mathbb{R}^*=\mathbb{R}P^2$ which would destroy spatial orientability. Requiring this condition, we restrict the antipode identification only to the $V,U$ coordinates.

\

$\cal{\bar{U}}$ and $\cal{W}$ dissapear, remaining $\cal{U}$, $\cal{B}$, $s_f$ $h_+$ and $h_-$. Along $h_-$, points denoted by $\sim$ are identified. The possibility of this identification was suggested by Rindler (Rindler, 1965), and also previously mentioned by Szekeres (Szekeres, 1960). The diagram in Fig. 2 is more related than that in Fig. 1 to the diagram corresponding to the final collapse of a spherical symmetric star and the formation of a real black hole, where $\cal{\bar{U}}$, $\cal{W}$, $s_p$ and $h_-$ do not exist. (This is the reason why the diagram in Fig. 1 is said to represent an ideal eternal black hole.) 

\

$SKS/2={{SKS}\over{\mathbb{Z}_2}}$ turns out to be a manifold with boundary
\begin{equation} 
\partial(SKS/2)=h_-\cong[0,+\infty)\times S^2_{2M}=Cy^3, 
\end{equation} 
an infinite 3-dimensional hypercylinder, and a conical singularity at $(V,U)=(0,0)$. For its topology one has 
\begin{equation}
SKS/2\cong(\mathbb{R}^2\times S^2)\cup Cy^3.
\end{equation}
The conical singularity must be taken off from the spacetime, resulting 
\begin{equation}
(SKS/2)^\prime\cong(\mathbb{R}^2\times S^2)\cup(Cy^3)^*
\end{equation}
with $h^\prime_-\cong h^\prime_+\cong(Cy^3)^*=\mathbb{R}^*\times S^2_{2M}$. (See Fig. 3.)

\begin{figure}[H]
	\centering
	\includegraphics[width=.5\linewidth]{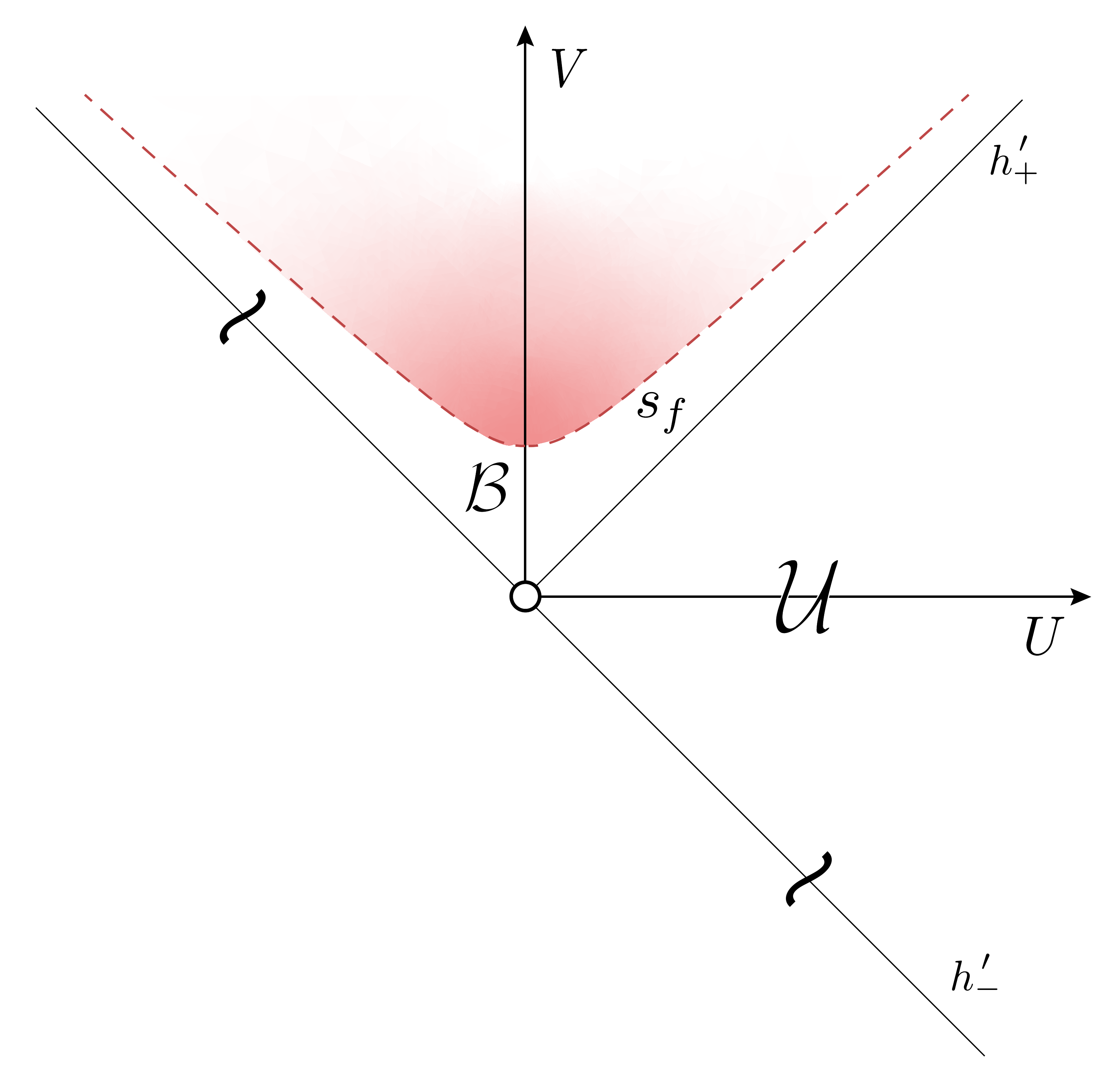}
	\caption{$(SKS/2)^\prime$}
	\label{fig:3}
\end{figure}

The dropping of the conical singularity seems to be done by hand; however, it is done to mantain differentiability at all points of the spacetime.
\section{Möbius transformation}
We can ask ourselves for another picture of the topology of $(SKS/2)^\prime$. With this aim, we consider the half plane ``above" $h^\prime_-$ as the complex half-plane
\begin{equation}
\mathbb{C}/2=\{z=x+iy, \ x\in\mathbb{R}, \ x\neq 0, \ y\in(0,+\infty)\}
\end{equation}
with the $x$-axis identified with $h^\prime_-$ ($z=x+i0$) (later we make the identification $x\sim -x$), and the half $y$-axis identified with $h^\prime_+$ ($z=iy$), and consider its image into the complex plane 
\begin{equation}
\mathbb{C}=\{w=\xi+i\eta, \ \xi,\eta\in\mathbb{R}\} 
\end{equation}
through the Möbius-type mapping 
\begin{equation}
\phi:(SKS/2)^\prime\to\mathbb{C}, \ z\mapsto\phi(z)=w:={{z-i}\over{z+i}}={{x+i(y-1)}\over{x+i(y+1)}}=\xi(x,y)+i\eta(x,y),
\end{equation}
and $S^2_r\mapsto S^2_r$, with 
\begin{equation}
\xi(x,y)={{x^2+y^2-1}\over{x^2+(y+1)^2}}, \ \eta(x,y)=-{{2x}\over{x^2+(y+1)^2}}.
\end{equation}
The relation between the $(V,U)$ coordinates and the $(y,x)$ coordinates is 
\begin{equation}
V={{y-x}\over{2}}, \ U={{y+x}\over{2}}.
\end{equation}
(The previous $\phi$ is a particular case of $z\mapsto\phi(z)=e^{i\gamma}({{z-z_0}\over{z-\bar{z}_0}}), \ \gamma\in\mathbb{R}, \ Im(z_0)>0$ (Churchill 1984), with $\gamma=0$ and $z_0=i$. $\phi$ maps the half plane $Im z=y>0$ onto the disk $\vert w\vert<1$, and the boundary of that half plane ($h^\prime_-$) onto the boundary of that disk.) 

\

The inverse of $\phi$ is given by 
\begin{equation}
z=z(w)=\phi^{-1}(w)=-i{{w+1}\over{w-1}}=x(\xi,\eta)+iy(\xi,\eta),
\end{equation}
with
\begin{equation}
x(\xi,\eta)=-{{2\eta}\over{(\xi-1)^2+\eta^2}}, \ y(\xi,\eta)=-{{(\xi^2-1)+\eta^2}\over{(\xi-1)^2+\eta^2}}.
\end{equation}
The Cauchy-Riemann ($C-R$) equations for $\phi$ and $\phi^{-1}$ are
\begin{equation}
{{\partial\xi}\over{\partial x}}={{\partial\eta}\over{\partial y}}, \ {{\partial\xi}\over{\partial y}}=-{{\partial\eta}\over{\partial x}}
\end{equation}
and
\begin{equation}
{{\partial y}\over{\partial\eta}}={{\partial x}\over{\partial\xi}}, \ {{\partial y}\over{\partial\xi}}=-{{\partial x}\over{\partial\eta}},
\end{equation}
respectively. The differentiability of $\phi$ and $\phi^{-1}$ guarantees that the coordinate transformation $(x,y)\to(\eta,\xi)$ is a genuine transformation in the context of general relativity.

\

$\phi$ is analytic: 
\begin{equation}
\phi^\prime(z)={{dw}\over{dz}}={{2i}\over{(z+i)^2}}
\end{equation}
and 1-1. Then, $(SKS/2)^\prime$ is onto its image and therefore homeomorphic and diffeomorphic to it, which turns out to be 
\begin{equation}
\phi((SKS/2)^\prime)=S^2\times(S^2_1\setminus(\{N,S\}\cup A)), \ A=\phi(s_f)\cup int(\phi(s_f)), 
\end{equation}
where $S^2_1$ is the unit 2-sphere, the identification $\sim$ is done, $S=\phi(0)=(\xi=-1,i\eta=0)$, and $N=(\xi=1,i\eta=0)=lim \ \phi(x+i0)$ as $x\to\pm\infty$. (See Fig. 4.) In turn, one has the homeomorphism
\begin{equation}
\phi((SKS/2)^\prime)\cong\mathbb{R}^{2*}\times S^2
\end{equation} 
since $S^2_1\setminus(\{S\}\cup A)\cong\mathbb{R}^2\setminus \{S\}\cong\mathbb{R}^{2*}$, with fundamental group $\pi_1(\mathbb{R}^{2*}\times S^2)\cong\pi_1(\mathbb{R}^{2*})\cong\mathbb{Z}$. Homotopically, then, 
\begin{equation}
(SKS/2)^\prime\cong\phi((SKS/2)^\prime)\simeq S^1\times S^2.
\end{equation}
Since $\phi$ is analytic and $\phi^\prime(z)\neq 0$, $\phi$ is also conformal, and so preserves the angles between tangents to intersecting curves. In particular this will be applied to the transformation of the light cones in $(SKS/2)^\prime$.

\begin{figure}[h!]
	\centering
	\includegraphics[width=\linewidth]{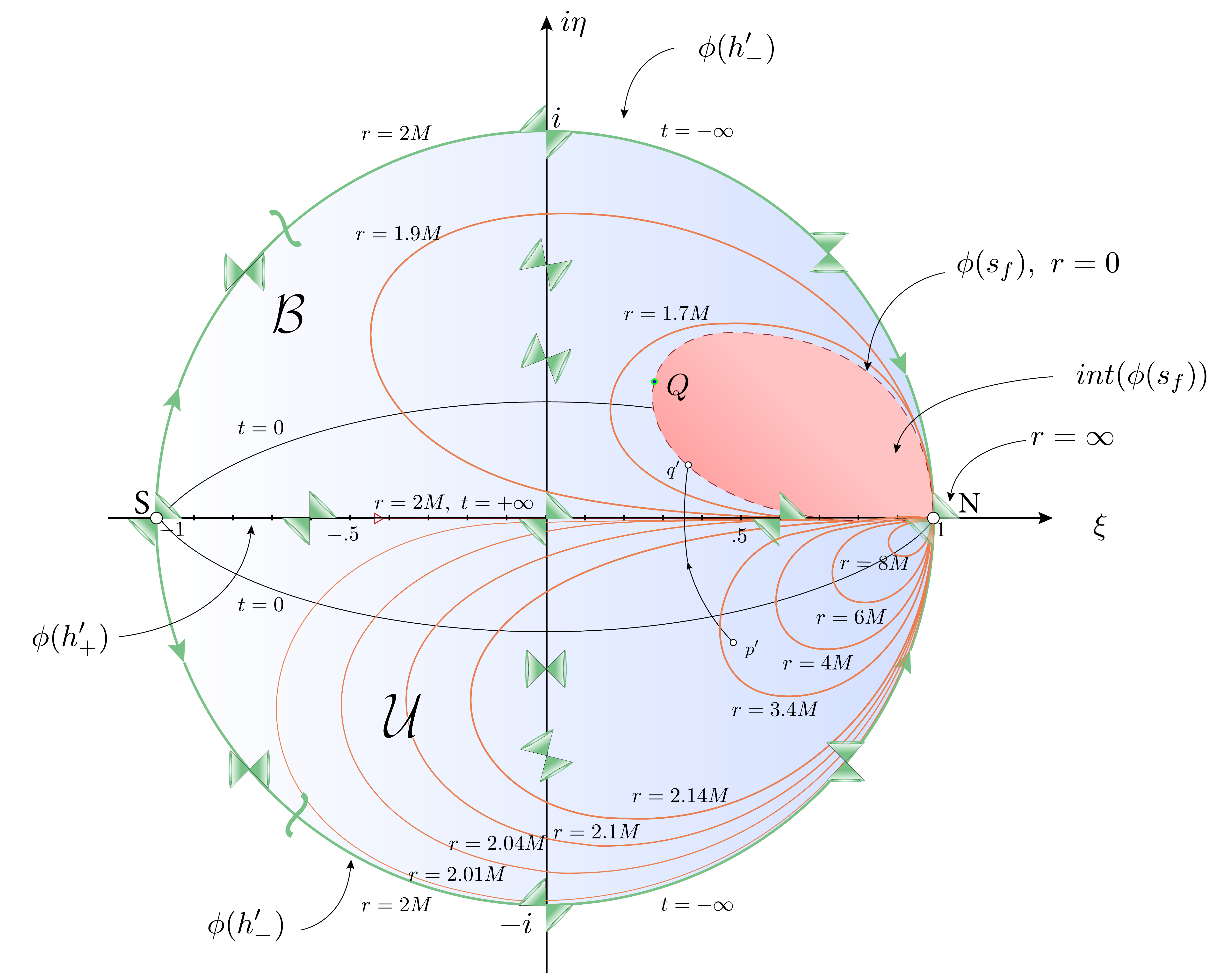}
	\caption{Image of $(SKS/2)^\prime$ under $\phi$; $p^\prime\to q^\prime$: null radial geodesic; $M=1$}
	\label{fig:4}
\end{figure}

The arrows in the image of $h^\prime_-$, 
\begin{equation}
\phi(h^\prime_-)=\partial(\phi((SKS/2)^\prime))=({{S^1_1\setminus\{N,S\}}\over{\sim}})\times S^2_{2M},
\end{equation}
corresponds to the path from $x=-\infty$ to $x=+\infty$ ($x\neq 0$) along $h^\prime_-$, and that in the image of $h^\prime_+$ corresponds to the path from $y=0_+$ to $y=+\infty$ along $h^\prime_+$. The image of $s_f$ is given by the red dashed curve, obtained from $\phi(z)$ with $y=-{{1}\over{x}}$, $x\in(-\infty,0)$; for the point $Q$ one has $Q=0.2+i0.4$. The fact that $S\notin\phi(h^\prime_+)$ implies that there is no (though infinite) closed causal (null) curve ($h^\prime_-$ in Fig. 3 or $\phi(h^\prime_-)$ in Fig. 4). Anyway, $h^\prime_-$ dissappears in a physical collapse. 
\section{$r$ and $t$ constant lines}
\subsection{$r$ lines}
i) $\cal{B}$ region:

\

For $0<r<2M$, $r=2\alpha M$ with $\alpha\in(0,1)$. From (4), $V^2-U^2=(1-\alpha)e^\alpha$ and using (13), $y=-{{(1-\alpha)e^\alpha}\over{x}}$, which replaced in (12) gives  
\begin{equation}
w_{\cal{B}}(x;\alpha)={{(x^4+(1-\alpha)^2e^{2\alpha}-x^2)+i(-2x^3)}\over{x^4+(1-\alpha)^2e^{2\alpha}-2x(1-\alpha)e^\alpha+x^2}}, \ x\in(-\infty,0).
\end{equation}
ii) $\cal{U}$ region:

\

For $2M<r$, $r=2\beta M$ with $\beta\in(1,+\infty)$. From (4), $y={{(\beta-1)e^\beta}\over{x}}$, which replaced in (12) gives
\begin{equation}
w_{\cal{U}}(x;\beta)={{(x^4+(1-\beta)^2e^{2\beta}-x^2)+i(-2x^3)}\over{x^4+(1-\beta)^2e^{2\beta}-2x(1-\beta)e^\beta+x^2}}, \ x\in(0,+\infty).
\end{equation}
For $\beta\to+\infty$, $y\to+\infty$, $w_{\cal{U}}(x;+\infty)=1+i0=N$. In Fig. 4 we plot some of these $r=const.$ lines.
\subsection{$t$ lines}
From (5), (13), and (15) we obtain
\begin{equation}
Th({{t}\over{4M}})={{\xi^2-1+\eta^2\mp2\eta}\over{\xi^2-1+\eta^2\pm2\eta}}
\end{equation}
with upper and lower signs respectively corresponding to the $r>2M$ ($\cal{U}$) and $r<2M$ ($\cal{B}$) regions. If we call 
\begin{equation}
\tau:=Th({{t}\over{4M}})\in(-1,+1), 
\end{equation}
it is easy to obtain $\eta_X(\xi;\tau)$ for $X=\cal{U}$ and $X=\cal{B}$:
\begin{equation}
\eta_{\cal{U}}(\xi;\tau)=({{1+\tau}\over{1-\tau}})-\sqrt{({{1+\tau}\over{1-\tau}})^2+(1-\xi)^2}, \ \eta_{\cal{B}}(\xi;\tau)=-({{1+\tau}\over{1-\tau}})+\sqrt{({{1+\tau}\over{1-\tau}})^2+(1-\xi)^2}.
\end{equation}
At the horizons $\phi(h^\prime_-)$ and $\phi(h^\prime_+)$, $t=-\infty$ and $t=+\infty$ respectively. For $t=0$ we have $\eta_{\cal{U}}(\xi;0)=1-\sqrt{2-\xi^2}$, $\eta_{\cal{B}}(\xi;0)=-1+\sqrt{2-\xi^2}$. These lines are plotted in Fig. 4.
\section{Transformed metric and light cones}
In the $(\xi,\eta)$-plane (remember that topologically $\mathbb{C}\cong\mathbb{R}^2$), the metric (3) becomes that of $\phi((SKS/2)^\prime)$, and is given by 
\begin{equation}
ds^2=-{{32M^3}\over{r}}e^{r/2M}(({{\partial x}\over{\partial\xi}})({{\partial x}\over{\partial\eta}})(d\eta^2-d\xi^2)-({{\partial x}\over{\partial\eta}}^2-{{\partial x}\over{\partial\xi}}^2)d\eta d\xi)-r^2(d\theta^2-sin^2\theta d\varphi^2)
\end{equation}
where the $C-R$ equations (16) and (17) were used, and
\begin{equation}
{{\partial x}\over{\partial\xi}}={{4(\xi-1)\eta}\over{((\xi-1)^2+\eta^2)^2}}, \ {{\partial x}\over{\partial\eta}}=-2{{(\xi-1)^2-\eta^2}\over{((\xi-1)^2+\eta^2)^2}}. 
\end{equation} 
$r=r(\xi,\eta)$ through the Lambert $W$ function and (13) and (15). (28) tells us that in the $(\xi,\eta)$ coordinates the light cones bend.

\

Given that $\phi(z)$ is analytic and, by (18), $\phi^\prime(z)\neq 0$, $\phi$ is conformal and therefore preserves angles between intersecting curves. Given two such curves in the $(V,U)$ plane (e.g. the lines corresponding to the boundary of the light cones), then their common rotation angle in the $(\xi,\eta)$ plane is given by the argument of $\phi^\prime(z)=\vert\phi^\prime(z)\vert e^{i\delta_{\phi^\prime(z)}}$:
\begin{equation}
\delta_{\phi^\prime(z)}=tg^{-1}({{x^2+(y+1)^2}\over{2x(y+1)}})=tg^{-1}({{1}\over{2}}({{1-\xi}\over{\eta}}-({{1-\xi}\over{\eta}})^{-1})).
\end{equation}
Some of these bended light cones are shown in Fig. 4.
\section{Null geodesics}
We analize here the image by $\phi$ of a typical radial null geodesic $p\to q$ in Fig. 1. In $SKS$, $p\to q$ is described by the equation $V=-U+2$ with $(V(p),U(p))=(0,2)$. It ``dies" at $s_f$ at the point $q$ where $-U+2=+\sqrt{1+U^2}$ which implies $U=3/4$ and  $V={5/4}$; so $(V(q),U(q))=(5/4,3/4)$. From (12) and (13),
\begin{equation}
\xi(V,U)={{2(V^2+U^2)-1}\over{2(V^2+U^2+V+U)+1}}, \ \eta(V,U)={{2(V-U)}\over{2(V^2+U^2+V+U)+1}};
\end{equation}
so the image of $p\to q$ in the $(\xi,\eta)$ plane is
\begin{equation}
\xi(U,-U+2)={{4(U^2-2U)-7}\over{4(U^2-2U)+13}}, \ \eta(U,-U+2)=-{{4(U-1)}\over{4(U^2-2U)+13}}.
\end{equation}
In particular, for $p^\prime=\phi(p)$, and $q^\prime=\phi(q)$, we obtain
\begin{equation}
(\xi(p^\prime),\eta(p^\prime))=(7/13,-4/13)\simeq(0.54,-0.31), \ (\xi(q^\prime),\eta(q^\prime))=(13/37,4/37)\simeq(0.35,0.11).  
\end{equation}
It is then easily verified that $q^\prime\in\phi(s_f)$ i.e. $\phi(p\to q)=p^\prime\to q^\prime$ dies at $\phi(s_f)$, as it must be. (See Fig. 4.) A similar analysis can be done with any other null geodesic in the image $\phi((SKS/2)^\prime)$.   
\section{Conclusion}
The antipodal identification $(V,U)\equiv(-V,-U)$ in the Schwarzschild-Kruskal-Szekeres ($SKS$) metric can be done without the introduction of additional singularities, since the requirement of differentiability makes it necessary to eliminate from the spacetime the emerging conical singularity. At the same time, this suppression guarantees the non existence of closed (though infinite) causal (null) curves. The Möbius transformation makes easier to study the topology of the resulting spacetime $(SKS/2)^\prime$ which, as expected, and in contradistinction with $SKS$, becomes non simply connected: $\phi((SKS/2)^\prime)\cong\mathbb{R}^{2*}\times S^2\simeq S^1\times S^2$, where $\simeq$ denotes homotopy type. The picture, however, of light cones, $r$ and $t$ constant lines, metric, and null geodesics, becomes much more involved than before the transformation.
\section*{Acknowledgments}
The author thanks Leonardo J. Méndez for numerical calculations, and Oscar Brauer for drawing the figures.
\section*{References}

\

Churchill, R.V. ``Complex Variables and Applications", 4th. edition (1984), pp. 194-195.

\

Kruskal, M.D. ``Maximal Extension of Schwarzschild Metric", Phys. Rev. {\bf 119} (1960) 1743-5.

\

Lambert, J.H. ``Observationes variae in mathesin purae", Acta Helveticae physico-mathematico-anatomico-botanico-medica, Band III (1758) 128-68. (Lambert W function, Wikipedia, 1-15.)

\

Rindler,W. ``Elliptic Kruskal-Schwarzschild Space", Phys. Rev. Lett. {\bf 15} (1965) 1001-2.

\

Sanchez, N. and Whiting, B.F. ``Quantum Field Theory and the Antipodal Identification of Black Holes", Nucl. Phys. B{\bf 283} (1987) 605-23.

\

Szekeres, Gy. ``On the Singularities of a Riemannian Manifold", Publicationes Mathematicae Debrecent {\bf 7} (1960) 285; reprinted: Gen. Rel. Grav. {\bf 34} (2002) 2001-16.

\

't Hooft, G. ``Virtual Black Holes and the Space-Time Structure", Found. Phys. {\bf 48} (2018) 1134-49.

\

\end{document}